\documentclass[aps,prb,twocolumn,superscriptaddress,showpacs]{revtex4}
\usepackage{amsfonts}
\usepackage{amssymb}
\usepackage{amsmath}
\usepackage{graphicx}
\usepackage{color}

\usepackage{dcolumn}   
\usepackage{bm}        
\usepackage{flafter}

\begin{document}


\title{Pressure effects on the superconducting transition in \textit{nH}-CaAlSi}

\date{\today}

\author{L. Boeri}
\affiliation{Max-Planck-Institut f\"{u}r
Festk\"{o}rperforschung, Heisenbergstra$\rm\beta$e 1, 70569
Stuttgart, Germany}
\author{J. S. Kim}
\affiliation{Max-Planck-Institut f\"{u}r Festk\"{o}rperforschung,
Heisenbergstra$\rm\beta$e 1, 70569 Stuttgart, Germany}
\author{M. Giantomassi}
\affiliation{ Unit\'e de Physico-Chimie et de Physique des Mat\'eriaux,
Universit\'e Catholique de Louvain, 1 place Croix du Sud, B-1348 Louvain-la-Neuve, Belgium}
\author{F. S. Razavi}
\affiliation{Department of Physics, Brock University, St.
Catharines, Ontario, L2S 3A1, Canada}
\author{S. Kuroiwa}
\affiliation{Department of Physics and Mathematics, Aoyama-Gakuin
University, Sagamihara, Kanagawa 229-8558, Japan}
\author{J. Akimitsu}
\affiliation{Department of Physics and Mathematics, Aoyama-Gakuin
University, Sagamihara, Kanagawa 229-8558, Japan}
\author{R. K. Kremer}
\affiliation{Max-Planck-Institut f\"{u}r Festk\"{o}rperforschung,
Heisenbergstra$\rm\beta$e 1, 70569 Stuttgart, Germany}

\begin{abstract}
We present a combined experimental and theoretical study of the
effects of pressure on $T_c$  of the hexagonal layered
superconductors \textit{nH}-CaAlSi (\textit{n} = 1, 5, 6), where
\textit{nH} labels the different stacking variants that were
recently discovered. Experimentally, the pressure dependence of
$T_c$ has been investigated by measuring  the magnetic
susceptibility of single crystals up to 10 kbar. In contrast to
previous results on polycrystalline samples, single crystals with
different stacking sequences
 display different pressure dependences of $T_c$.
$1H$-CaAlSi shows a decrease of $T_c$ with pressure, whereas $5H$
and $6H$-CaAlSi exhibit an increase of $T_c$ with pressure.
\textit{Ab-initio} calculations for $1H$, $5H$ and $6H$ -CaAlSi
reveal that an ultrasoft phonon branch associated to  out-of-plane
vibrations of the Al-Si layers softens with pressure, leading to a
structural instability at high pressures. For $1H$-CaAlSi  the
softening is not sufficient to cause an increase of $T_c$, which
is consistent with the present experiments, but adverse to
previous reports. For $5H$ and $6H$ the softening provides the
mechanism to understand the observed increase of $T_c$ with
pressure. Calculations for hypothetical $2H$ and $3H$ stacking
variants reveal qualitative and quantitative differences.
\end{abstract}
\smallskip


\maketitle

Superconductivity in hexagonal layered compounds has attracted
broad interest since the discovery of "high-$T_c$"
superconductivity in MgB$_2$~\cite{MgB2:nagamatsu:syn,MgB2:PhysicaC} and other
structurally related compounds such as
(Ca,Sr)AlSi,\cite{CaAlSi:imai:syn}
CaSi$_2$,\cite{CaSi2:sanfilippo:Tc} and very recently,
alkaline-earth intercalated
graphites.\cite{YbC6:weller:syn,CaC6:emery:syn,SrC6:kim:syn} In all
these compounds the light elements are arranged in honeycomb
layers which are intercalated by alkaline earth atoms. Depending
on the elements forming the honeycomb layer, the electron and
phonon states involved in the superconducting pairing are quite
different. In MgB$_2$, holes in the $\sigma$ bands of the B layer
couple strongly to B bond-stretching phonon modes.
 In CaSi$_2$,
 which can be considered as the "antibonding analogue"
to MgB$_2$, the $\sigma^*$ bands are strongly coupled to the
bond-stretching phonons of the Si layer. In CaC$_6$ and CaAlSi,
the so-called "interlayer" bands are filled and experience
significant electron-phonon ($e$-ph) interaction to the
out-of-plane buckling vibrations of the honeycomb
layers.\cite{AAlSi:mazin:band,CaAlSi:giantomassi:band,CaC6:calandra:band,CaC6:jskim:pressure,GIC:boeri:band}

CaAlSi is of particular interest because it  exhibits an ultrasoft
phonon mode and  crystalizes with several stacking
variants.\cite{CaAlSi:sagayama:Xray,CaAlSi:kuroiwa:syn} The
interplay of these two ingredients gives rise to intriguing
effects both on the normal and the superconducting properties. The
presence of an ultra-soft phonon  mode at $\sim$7 meV was
initially evidenced by first-principles
calculations\cite{AAlSi:mazin:band,AAlSi:huang:band,AAlSi:huang:pressure,CaAlSi:giantomassi:band,AAlSi:heid:band}
and recently confirmed by neutron-scattering
experiments.\cite{AAlSi:heid:band} The ultra-soft phonon modes are
believed to induce strong $e$-ph coupling and to lead to an
enhanced specific heat anomaly at $T_c$ as well as to  a $positive$
pressure dependence of $T_c$ in CaAlSi. These effects have not
been observed in the isoelectronic and isostructural SrAlSi which
does not display any signature of soft phonon modes.
\cite{AAlSi:lorenz:pressure}

\begin{figure}[htb]
\begin{center}
\includegraphics[width=8.0cm,bb=35 460 460 775]{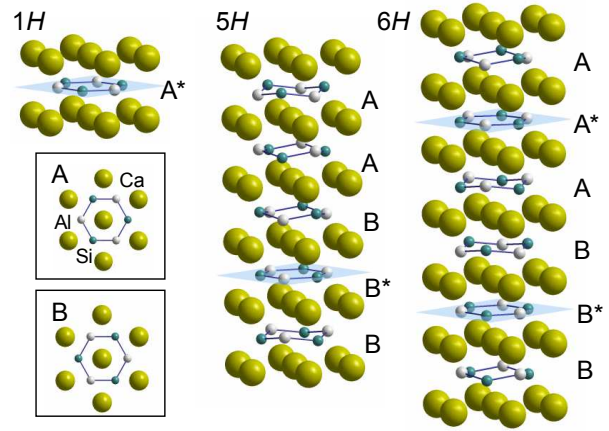}
\end{center}
\caption{\label{fig:structure}(color online) Crystal structure of
$1H$, $5H$ and $6H$- CaAlSi,
 which are characterized by a different sequence of A-
and B-layers. A- and B-layers are rotated by 60$^{\rm o}$ around the
$c$-axis with respect to each other. Flat and buckled Al-Si layers
 are indicated with and without an asterisk,
respectively.}
\end{figure}
Recent X-ray diffraction experiments on \textit{nH}-CaAlSi single
crystals revealed several stacking variations of
two different types of  Al-Si
layers, denoted as A and B in the following (see Fig.~\ref{fig:structure}). 
The A- and B-layers differ
by a 60$^{\rm o}$ rotation around the $c$-axis.
\cite{CaAlSi:sagayama:Xray,CaAlSi:kuroiwa:syn} Besides the  simple
$1H$ structure characterized by a $|A|A|A|...$ stacking, two more
stacking variants were found: $5H$ with a $AABBB$ and $6H$
with a $AAABBB$ sequence. Stacking of the $A$ and $B$ layers induces an
internal stress on the structure, causing a buckling of
\textit{boundary layers}, i.e. layers with an unlike neighboring
layer. Subsequent investigations show that the superconducting
properties strongly depend on the kind of stacking of A- and
B-layers.\cite{CaAlSi:kuroiwa:syn,CaAlSi:kuroiwa:tunnelling,CaAlSi:prozorv:microwave,CaAlSi:lupi:IR}
However, all \emph{ab-initio} calculations, and analyses of the
experimental data based thereon, so far have assumed that the
stacking of Al-Si planes along the $c$ axis is either uniform, or
completely disordered. To clarify the complex interplay of
stacking variants, buckling, soft modes and superconductivity, further
investigation of the $e$-ph properties of different stacking
variants of CaAlSi are highly desirable.

In this paper, we investigate the effects of pressure on the
superconducting properties of \textit{nH}-CaAlSi by experiments on
single crystals and \textit{ab-initio} calculations of the $e$-ph
properties. Previous experiments on polycrystalline samples of
CaAlSi and SrAlSi have shown that $T_c$ of CaAlSi increases under
pressure while $T_c$ of SrAlSi decreases.
\cite{AAlSi:lorenz:pressure} In an {\em ab-initio} study, Huang
\textit{et al.} have proposed that in $1H$-CaAlSi an enhancement
of the $e$-ph coupling associated to the ultra-soft phonon modes
can overcome the negative contribution of the other phonon
branches and lead to an increase of $T_c$ under pressure. Such an
ultra-soft phonon mode was not obtained for
SrAlSi.\cite{AAlSi:huang:pressure}

However,   recent experimental and theoretical results
question these findings. First, until now the pressure dependence  of $T_c$
in polycrystalline samples has been interpreted assuming a
$1H$ stacking variant. However,
 we will demonstrate here
that the pressure dependence of $T_c$  of $1H$-CaAlSi single
crystals  is markedly different from that of polycrystalline
samples.
Secondly, in these calculations\cite{AAlSi:huang:pressure},
 the soft phonon branch in CaAlSi is
unstable in some part of  the brillouin zone (BZ),
already at ambient pressure. This instability was not confirmed by
recent \textit{ab-initio} calculations and neutron scattering
experiments.\cite{CaAlSi:giantomassi:band,AAlSi:heid:band} Finally,
Huang \textit{et al.} assumed a uniform
compressibility, which is very unlikely to occur in layered
materials.

The aim of our analysis is two-fold: on the one hand we wish to
investigate the interplay between soft modes and superconductivity
in  $1H$-CaAlSi based on \textit{ab-initio} calculations
and experiments on single crystals. On the other
hand, we wish to understand how stacking variants affect the
pressure dependence of the superconducting properties of CaAlSi.

Single crystals of CaAlSi were grown by the floating zone method.
Details of the crystal growth and characterization are described
elsewhere.\cite{CaAlSi:kuroiwa:syn} The superconducting transition
temperature ($T_c$) under pressure was determined  by measuring the
magnetic susceptibility { in} a SQUID magnetometer
(Quantum Design). A Cu-Be piston-anvil-type pressure cell was used
to apply quasi-hydrostatic pressures up to $P$ $\sim$ 10 kbar with
silicon oil or Fluorinert as pressure transmitting medium. To
monitor the pressure inside the pressure cell, we
performed $in$-$situ$ measurements of the $T_c$ of Sn (purity 99.999$\%$)
or Pb (purity 99.9999 $\%$).~\cite{Schilling}

\begin{figure}[bth]
\includegraphics[width=8.2cm,bb=10 37 250 260]{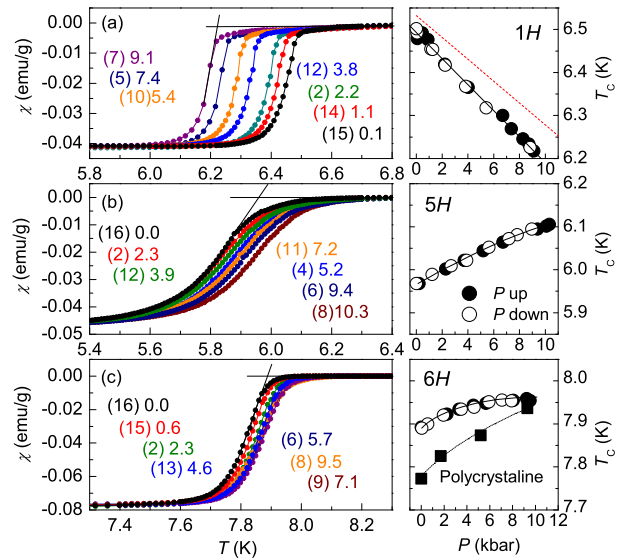}
\caption{\label{fig:suscept} (color online) \textit{Left:}
Temperature dependence of the susceptibility for CaAlSi single
crystals at different pressures. The numbers next to the data and
in the parenthesis correspond to the applied pressure (kbar) and
the sequential order of the measurement runs, respectively. The
extrapolation method to determine $T_c$ is demonstrated with solid
lines. \textit{Right:} Pressure dependence of $T_c$ for (a) $1H$-,
(b) $5H$-, (b) $6H$-CaAlSi single crystals. For comparison we plot
the previous data for polycrystaline sample from
(Ref.~\onlinecite{AAlSi:lorenz:pressure}) in (c) and the
theoretical curve for $T_c(P)$ in (a) for $1H$-CaAlSi with (red)
dashed line.~\cite{footnote:plot}}
\end{figure}

Figure~\ref{fig:suscept} shows the temperature dependence of the
magnetic susceptibility ($\chi(T)$) of single crystals of
\textit{nH}-CaAlSi under pressures up to $\sim$ 10 kbar. At
ambient pressure, the $T_c$'s of $1H$, $5H$, and $6H$-CaAlSi are
6.50 K, 5.95 K, and 7.89 K, respectively, consistent with previous
reports.\cite{CaAlSi:kuroiwa:syn}
 The superconducting transitions are relatively sharp for
$1H$ and $6H$-CaAlSi with a transition width of $\Delta$$T_c$ $\sim$
0.1 K at 80$\%$ of the diamagnetic shielding, indicating
 a  good sample quality. For $5H$-CaAlSi, the
 transition is broader ($\Delta T_c$ $\sim$ 0.4 K),
but still sharp enough to follow its pressure dependence. With
increasing pressure, the superconducting transition for $1H$-CaAlSi
clearly shifts to lower temperatures without significant broadening
of the transition. In contrast,  $T_c$ in $5H$ and
$6H$-CaAlSi increases with pressure. For all
samples, the $T_c$'s and the shape of $\chi(T)$ trace taken after
gradually releasing again the pressure agreed with the data
collected with increasing pressure.

Our measurements reveal that the pressure variation of $T_c$
of $nH$-CaAlSi
depends strongly on the stacking sequence. For the detailed
comparison, $T_c$ was identified as the temperature where the
extrapolation of the steepest slope of $\chi(T)$ intersects the
normal-state susceptibility extrapolated to lower temperatures.
Using different criteria to determine $T_c$, e.g. the mid-point of
the transition, changes the absolute value of $T_c$, but does not
alter the relative variation of $T_c$. Surprisingly, for $1H$-CaAlSi,
$T_c$ decreases linearly with a rate of $\Delta T_c$/$T_c$ = -0.03
K/kbar, in contrast to a previous report on polycrystalline
samples.\cite{AAlSi:lorenz:pressure} For $5H$-CaAlSi the pressure
dependence is slightly nonlinear with an initial slope of +0.013
K/kbar, while $6H$-CaAlSi exhibits a more pronounced nonlinear
behavior and saturation at $T_c$ $\sim$ 7.95 K already at $\sim$
10 Kbar.

In order to understand the effects of  pressure and
 stacking variants  on the superconducting properties of CaAlSi,
we performed \textit{ab-initio} (DFPT)
calculations~\cite{DFT:baroni:RMP,PWSCF,ABINIT} of the electronic
and vibrational properties of \textit{nH}-CaAlSi as a
function of pressure.  The computational details are the same as in
Ref.~\onlinecite{CaAlSi:giantomassi:band}, unless otherwise
specified.~\cite{note:cal}
Besides the identified 1$H$, 5$H$ and 6$H$ stacking variants, we
also considered hypothetical
$2H$ and $3H$-CaAlSi,
characterized by an $|AB|AB|$ and $|AAB|AAB|$ stacking of the Al-Si
planes, respectively.

In the investigated pressure range up to 100 kbar, we
relaxed the internal parameters of all structures at
 intervals of 20 kbar. The details of the
optimization, together with the detailed band structures and phonon
dispersion relations, will be published elsewhere. We subsequently
fitted the $E$ vs. $V$ curves to a Birch-Murnagham
equation-of-state, to obtain the equilibrium volume $V_0$ and the bulk
modulus $B_0$ for each system.
All equilibrium volumes and bulk moduli are close to each other
($68.6~\le~V_0~\le~69.0~\rm{\AA}^3$, $49.6~\le~B_0~\le~51.8$~GPa).
In sign and magnitude, our calculations correctly reproduce  the
corrugation (``buckling") of the Al-Si planes in the $5H$ and $6H$
structures. A corrugation is also obtained for $3H$-CaAlSi. In
contrast to previous suggestions,\cite{CaAlSi:sagayama:Xray} the
corrugation in the Al-Si layers of the $5H$ and $6H$ phases is not
reduced by applying pressure. Rather, it remains unchanged for
$6H$, or becomes even more pronounced for $5H$-CaAlSi. The
compressibility is anisotropic, $k_c/k_a =3.8$, as expected for a
layered material. At all pressures, the total energy of
$2H$-CaAlSi and $3H$-CaAlSi is higher than in $1H$, $5H$ and
$6H$-CaAlSi, the latter  being almost degenerate in energy.~\cite{footnote:energy}
\begin{figure}
\includegraphics[width=8.5cm,bb=10 15 200 155]{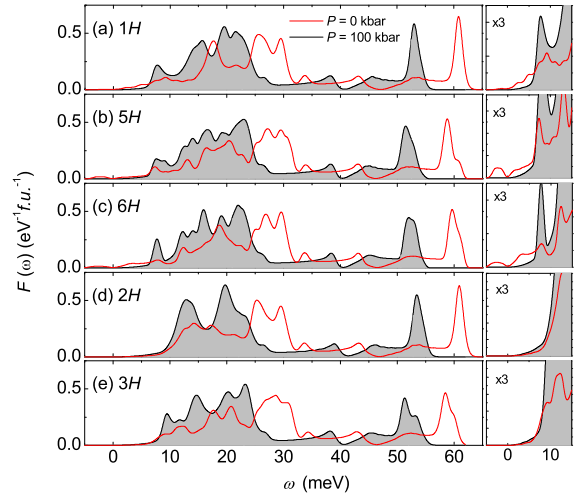}
\caption{\label{fig:phDOS} (color online) Linear Response Phonon
DOS (phDOS) for different $nH$-CaAlSi at $P$=0 and $P$=100 kbar;
in the right sub-panel we show an enlargement of the low-energy
region.}
\end{figure}

Using the relaxed structures, we calculated by linear response
 the phonon dispersion relations
{ $\omega_{\mathbf{q},\nu}$} and the $e$-ph linewidths
$\gamma_{\mathbf{q},\nu}$ on a regular grid in $\mathbf{q}-$space.
The phonon dispersions and the phonon Densities of States
(phDOS) were then obtained by Fourier interpolation of
the dynamical matrices, while the Eliashberg spectral functions
$\alpha^2 F(\omega)$ and the total $e$-ph coupling
$\lambda$'s were calculated by summing over individual linewidths
and frequencies.

In Fig.~\ref{fig:phDOS}, we display the phDOS of $nH$-CaAlSi,
calculated at the theoretical equilibrium pressure ($P$ = 0) and
at the highest pressure considered (100 Kbar). At $P$ = 0, the
1$H$, 5$H$ and 6$H$ stacking variants all show a
 peak at $\sim$ 7 meV, associated with the
out-of-plane vibrations of the Al-Si planes. Under
pressure, roughly one half of the corresponding phonon states
 soften and drive the
system to a structural instability at $P \sim 80$ Kbar.~\cite{note}
The hypothetical $2H$-CaAlSi and $3H$-CaAlSi, on the other hand, do
not display the ultra-soft phonon peak at $P$ = 0, and all phonon
modes harden with increasing pressure.

At present, we cannot conclusively explain the different phonon
softening behavior of the systems. However, by comparing our
(partial) phDOS's obtained for different $nH$ stacking variants
with those of previous VCA
calculations,~\cite{CaAlSi:giantomassi:band,AAlSi:heid:band} we
observe that phonon softening mainly involves Al out-of-plane
vibrations.
 Phonon softening appears if a
block of three or more like planes is found in the sequence.
This probably reflects the tendency of Al to form bonds with
other Al atoms in the neighbouring layers
rather than $sp^2$ bonds with Si in the same layer,
if the $c$ lattice constant is small enough.

 Our results strongly differ from those of
Ref.~\onlinecite{AAlSi:huang:pressure} for $1H-$CaAlSi. At $P=0$,
our phonon frequencies are all real, and even at higher pressures
only a few become imaginary, while in
Ref.~\onlinecite{AAlSi:huang:pressure} the whole out-of-plane
branch becomes unstable. The most likely reason for this
discrepancy is that we optimized the structural parameters, while
the previous work assumed an isotropic compressibility. Also, a
careful convergence of phonon frequencies with respect to
$\mathbf{k}$-sampling and basis set size is crucial to obtain the
correct behaviour of the ultra-soft phonon modes.

\begin{figure}
\includegraphics[width=8.0cm,bb=10 15 195 95]{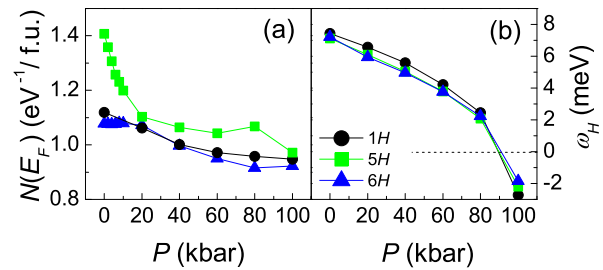}
\caption{\label{fig:DOS_wH} (color online) (a) Pressure dependence
of the DOS at  $E_F$  and (b) frequency of the lowest-lying phonon
branches at the $H$ point  for various
$nH$-CaAlSi.~\cite{note:caption}}
\end{figure}

If the coupling to the electrons is strong
enough to overcome the  opposing effect of the
hardening of the other phonons and of the decrease of the electronic
DOS at the Fermi level,  a soft phonon mode
 can lead to an increase of $T_c$ with
pressure.
 In Fig.~\ref{fig:DOS_wH} we show the calculated
pressure dependences of the electronic DOS at the Fermi level,
$N(E_F)$, and of the lowest lying phonon frequency at the $H$
point ($\omega_H$) for $1H$, $5H$ and $6H$-CaAlSi. We estimate the
pressure evolution of the partial $e$-ph coupling constant associated to
the soft phonon mode, using the Hopfield formula:
$\lambda=N(E_F)I^2/\omega^2_{H}$, where $I$ is the $e$-ph matrix
element. Assuming that $I$ is independent of pressure, we estimate
that the partial $\lambda$ increases by $\sim$ 60$\%$ for $1H$ and
$6H$, and $\sim$ 50$\%$ for $5H$. For $2H$ and $3H$-CaAlSi (not
shown), $N(E_F)$ decreases and $\omega_{H}$ increases, resulting
in a net decrease of $\lambda$. From this argument, one could
expect a very similar (increasing) behavior of $T_c$ with pressure
for $n=1,5, 6$. This is, however, not observed experimentally.
\begin{figure}
\includegraphics*[width=8.0cm,bb=10 10 240 160]{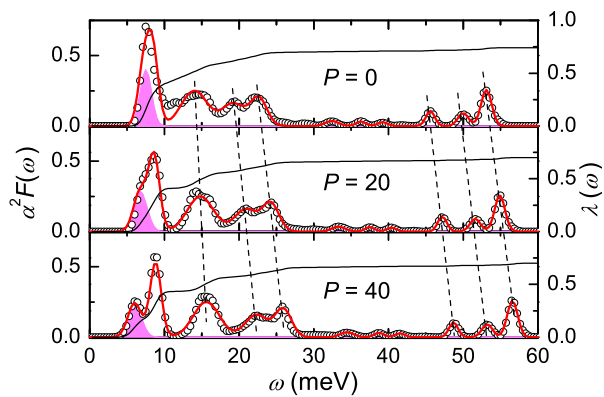}
\caption{\label{fig:alpha}(color online ) Evolution of the LDA
Eliashberg function (open circles) $\alpha^2 F(\omega)$ and
$\lambda(\omega)$ with pressure (kbar), for 1$H$-CaAlSi.
 Red solid lines
correspond to the simplified model with Gaussian peaks, described
in the text. The resulting $\left<\omega_{ln}\right>$ and
$\lambda$ are listed in Table.~\ref{table:table1}. The shaded area
highlights the  variation of energy and intensity for the
ultra-soft phonon modes. }
\end{figure}

Figure~\ref{fig:alpha} shows the evolution of the Eliashberg
function $\alpha^2 F(\omega)$ of $1H$-CaAlSi for which we ran a
full $e$-ph calculation as a function of pressure. In agreement
with previous
calculations,\cite{CaAlSi:giantomassi:band,AAlSi:heid:band} at $P$
= 0 there is a large peak at $\omega \sim 7$ meV, associated to
the Al-Si out-of-plane vibrations in the $k_z$=$\pi/c$ plane,
implying that the corresponding phonon states have a considerable
coupling to electrons. At higher pressures, following the behavior
of the phDOS, this peak splits into two peaks, one of them
softening, the other hardening with increasing pressure.
The pressure-induced transfer of the spectral weight
from the hard to the soft part of the low-lying peak  affects both
the total $e$-ph coupling parameter
$\lambda=2\int_{0}^{\infty}\alpha^2 F(\omega)\omega^{-1} d\omega$
and the logarithmic-averaged phonon frequency
$\langle\omega_{\ln}\rangle$. Using the Allen-Dynes
equation:~\cite{AllenDynes}
\begin{equation}
T_c =\frac{\langle\omega_{ln}\rangle}{1.2} \exp \left[
\frac{-1.04(1+\lambda)}{\lambda-(1+0.62\lambda)\mu^*} \right],
\label{eq:allen}
\end{equation}
where $\mu^*$ is the Coulomb pseudopotential, fixed to $\mu^*=0.1$
in the following (see Tab.~\ref{table:table1}), we obtain a  {\em
decrease} of $T_c$ with pressure for $1H$-CaAlSi, in agreement with
our experiments. A slight variation of $\mu^*$ does
not alter this general behavior.

In stark contrast to the previous report for
1$H$-CaAlSi,\cite{AAlSi:huang:pressure} the effect of softening
of the ultra-soft phonon modes is apparently not sufficient to overcome the
opposing contributions from the other phonon branches. For further
model calculations, we used a decomposition of $\alpha^2F(\omega)$
of $1H$-CaAlSi into Gaussians as shown in Fig.~\ref{fig:alpha}.
Using this decomposition reveals that the behavior of $T_c$ with pressure
follows that of the spectral weight of the ultra-soft phonon mode,
which is decreasing with pressure in $1H$-CaAlSi. The change of
the spectral weight of the ultra-soft phonon mode determines the
behavior of $T_c$ under pressure, more importantly than the
decrease of its phonon frequency.

For  $5H$ and $6H$-CaAlSi, due to their very large unit cells
(15 and 18 atoms, respectively), and  the extreme sensitivity
of the results to computational parameters, we did not perform a
full $e$-ph calculation.
 Although there are some subtle differences in the
electron and phonon dispersions, the results shown in
Fig.~\ref{fig:phDOS} and \ref{fig:DOS_wH} imply that the  pressure
dependence is qualitatively very close to that of $1H$-CaAlSi, ({\em
i.e.} the $N(E_F)$'s and $\omega_H$ decrease with the same rate,
and a structural instability happens at elevated pressures).
 For this reason, we
assume that their $\alpha^2 F(\omega)$'s have very similar
characteristics to that of $1H$-CaAlSi.  The different pressure
dependence  of $T_c$ for $5H$ and $6H$  therefore has to be
attributed to a different (increasing) behavior of the spectral
weight for the ultra-soft phonon mode under pressure. This
behavior could reflect either small differences in $e$-ph matrix
elements due to the buckling of some planes, or a different number
of phonon modes that soften under pressure.

\begin{table}[b]
\begin{ruledtabular}
\begin{tabular}{lccc}
   $P$ (Kbar) & $\left<\omega_{ln}\right>$ (K)   & $\lambda$           & $T_c$ (K)            \\
\hline
        0   &     139.2   (160)   &    0.73  (0.60)       &     5.35    (3.66)    \\
       20   &     139.9   (186)   &    0.70  (0.47)       &     4.86    (1.75)    \\
       40   &     134.0   (200)   &    0.70  (0.43)       &     4.65    (1.24)    \\
\end{tabular}
\end{ruledtabular}
\caption{\label{table:table1}Calculated superconducting properties
of $1H$-CaAlSi as a function of pressure. $T_c$ was obtained by
Allen-Dynes formula, with $\mu^*=0.1$. For comparison, we give in
parentheses the results for $2H$. }
\end{table}

Another aspect, which we cannot rule out completely is, that in
$5H$ or $6H$-CaAlSi  multi-band effects become relevant. Recently
Lupi \textit{et al.} reported an anisotropy of the optical
response in the superconducting and the normal states
for a crystal with mixed 5$H$ and 6$H$ phases.\cite{CaAlSi:lupi:IR} The
superconducting gap, determined by penetration depth measurements,
also shows a sizable anisotropy, depending on the stacking
sequences.\cite{CaAlSi:prozorv:microwave} Further studies are
needed to clarify this point. In particular,
understanding the effect of buckling of the Al-Si layers
 on the anisotropy of the $e$-ph coupling appears to be
the crucial issue.

We would like to emphasize once more that the hypothetical systems 
$2H$ and $3H$-CaAlSi show
a completely different behavior.
First, they are energetically disfavored with respect to existing
stacking variants, because of a non-optimized energy balance between
the formation of AB interfaces and buckling.
 Second, they do not display
any soft phonon modes,
 which only appear if three or more Al atoms arrange in a sequence
along the $c$-axis.

 All  phonon modes harden with
pressure, leading to an increase in $\left<\omega_{ln}\right>$,
 a fast decrease of $\lambda$, and a net decrease of
$T_c$ with pressure with a rate of $dT_c/dP$ $\sim$ -0.05 K/Kbar.
For comparison, the results for $2H$-CaAlSi are also listed in Table
~\ref{table:table1}.   
$2H$ and $3H$-CaAlSi, therefore, do not
represent proper models for other stacking variants.

In summary, we demonstrate that for single crystals of
\textit{nH}-CaAlSi the behavior of $T_c$ under pressure depends
crucially on the particular stacking sequence and the buckling of
the Al-Si layers. $T_c$ {\em decreases} for $1H$, and {\em
increases} with pressure in $5H$ and $6H$-CaAlSi. Previous
experiments on polycrystaline samples gave results that are closer
to the behavior of $6H$-CaAlSi, but adverse to the behavior of
$1H$-CaAlSi. One may speculate that the polycrystalline sample
consisted mainly of phases with stacking variants other than
1$H$-CaAlSi. Based on our \textit{ab-initio} calculations we find
a gradual softening of an out-of-plane phonon mode under pressure
for \textit{nH}-CaAlSi (\textit{n} =1, 5, 6), which leads to a
structural instability at higher pressures. In $1H-$CaAlSi, the
softening is not strong enough to lead to an increase in $T_c$, in
contrast to previous calculations, while it is likely that this
softening leads to the increase of $T_c$ with pressure in single
crystals of $5H$ and $6H$-CaAlSi.

\acknowledgments The authors acknowledge useful discussion with K.
Syassen, A. Simon, O. K. Andersen and G. B. Bachelet. 
We also thank E. Br\"{u}cher,
S. H\"{o}hn for experimental assistance.


\begin{thebibliography}{}
\bibitem{footnote:plot} In panel ($a$), to obtain the theoretical $T_c$ we used the Allen-Dynes formula
with  $\mu^{*}=0.072$ in order to match the experimental $T_c$ at $P$=0. The
other parameters are in table \ref{table:table1}. 
In panel ($c$), the $T_c$ vs. $P$ curve
for polycrystalline samples is obtained from 
 the raw susceptibility data of Ref.~\onlinecite{AAlSi:lorenz:pressure},
using the same criteria for determining $T_c$ that we used for single crystals.
\bibitem{note:cal} In the present paper, we devoted special care to the
convergence of the $e$-ph coupling functions with pressure, which
turned out to depend critically on the $\mathbf{q}$-space
integration, particularly along the $c$ direction. For the higher
pressures, a $ 6\times6\times12$ $\mathbf{q}$-grid was found
necessary to ensure convergence.
\bibitem{note:caption}
For consistency, we used the theoretical lattice constants at all
pressures. At $P=0$, the deviation between the theoretical and
experimental lattice constant is less than $1 \%$, which has no
effect on the calculated phonon frequencies for all systems and on
the electronic DOS for 1$H$ and 6$H$. For 5$H$, using the
experimental lattice constant at $P = 0$ we obtain a much lower
value for $N(E_F)$ = 1.1 st/eV f.u., in perfect agreement with
other calculations (Ref.~\onlinecite{AAlSi:Kuroiwa:band}), and
much closer to that of 1$H$ and 6$H$.
\end{thebibliography}
\end{document}